\documentclass[conference]{IEEEtran}
\usepackage{times,amsmath,epsfig,cite}
\title{Design and Implementation of BCM Rule \\Based on Spike-Timing Dependent Plasticity}

\author{%
{Mostafa Rahimi Azghadi, Said Al-Sarawi, Nicolangelo Iannella, and Derek Abbott}
\vspace{1.6mm}\\
\fontsize{10}{10}\selectfont\itshape
Centre for Biomedical Engineering, School of Electrical and Electronic Engineering,\\
The University of Adelaide, Adelaide, SA 5005, Australia\\\fontsize{9}{9}\selectfont\ttfamily\upshape
\{mostafa,alsarawi,iannella,dabbott\}@eleceng.adelaide.edu.au
}
\begin{document}
\maketitle
\begin{abstract}
The Bienenstock-Cooper-Munro (BCM) and Spike Timing-Dependent Plasticity (STDP) rules are two experimentally verified form of synaptic plasticity where the alteration of synaptic weight depends upon the rate and the timing of pre- and post-synaptic firing of action potentials, respectively. Previous studies have reported that under specific conditions, i.e. when a random train of Poissonian distributed spikes are used as inputs, and weight changes occur according to STDP, it has been shown that the BCM rule is an  emergent property. Here, the applied STDP rule can be either classical pair-based STDP rule, or the more powerful triplet-based STDP rule. 
In this paper, we demonstrate the use of two distinct VLSI circuit implementations of STDP to examine whether BCM learning is an emergent property of STDP. These circuits are stimulated with random Poissonian spike trains. The first circuit implements the classical pair-based STDP, while the second circuit realizes a previously described triplet-based STDP rule. 
These two circuits are simulated using \boldmath{$0.35$}~\boldmath{$\mu$}m CMOS standard model in HSpice simulator. Simulation results demonstrate that the proposed triplet-based STDP circuit significantly produces the threshold-based behaviour of the BCM. Also, the results testify to similar behaviour for the VLSI circuit for pair-based STDP in generating the BCM.
\end{abstract}



\section{Introduction}
Synaptic plasticity is the study of how synapses (and the connection between neurons) change their functional or structural properties in an activity-dependent manner. This change is believed to be due to some set of underlying molecular processes and has been intensively studied in the last two decades \cite{bear1986,clothiaux1991,kirkwood1993,kirkwood1996,lisman2005,quinlan1999,ref1,ref2,ref3,ref4,ref5,sjostrom2008}. Functional changes in synapses typically give rise to alterations (increase or decrease) in the number of  $\alpha$-amino-3-hydroxy-5-methyl-4-isoxazolepropionic acid (AMPA) receptors. Attempts to understand how such changes influence network behaviour required the construction of quantitative, but simplified mathematical models, which capture the essential nature of such plasticity processes. Understanding how these underlying molecular processes give rise to learning is believed to be one of the most important pieces of the neural learning and memory puzzles. There are two major classes of biologically inspired models for synaptic plasticity. One class only considers the timing difference between pre- and post-synaptic action potentials~\cite{ref1,ref6,ref7}, while the other considers rate of pre-synaptic inputs and post-synaptic action potentials~\cite{ref8,ref9}. These classes of model have mainly been studied from the theoretical/computational point of view and use various degrees of model complexity, ranging from simple models using integrate-and-fire neurons \cite{ref5} through to investigations studying the impact of synaptic plasticity, including STDP, on the spatial arrangement of synaptic inputs on the dendrites of spatially extended neurons~\cite{iannella2006,iannella2010,gidon2009,ilan2011interregional}. Significantly, some researchers have been physically realizing them as VLSI circuits to provide learning components for physical neural networks~\cite{ref10,ref11,ref12,ref13,ref14,rachmuth2011biophysically}.

In addition to these major works, a Spike-Driven Synaptic Plasticity (SDSP) rule was developed to address the shortcomings of the pair-based STDP rule to faithfully learn patterns of mean firing rate. The SDSP rule employs both the timing of the pre-synaptic and the rate of the post-synaptic action potentials to induce synaptic weight changes~\cite{brader2007learning,ref11}. This rule can be considered as a hybrid rule which does not fall in either of the classes mentioned above. Here, we show that triplet-based STDP can give rise to the BCM rule which is a rate-based rule. In addition, according to a very recent study~\cite{gjorgjieva2011triplet}, triplet-based STDP can generalize the BCM rule in a way that it can induce selectivity with rate-based patterns with mean firing rates. Therefore, contrary to pair-based STDP, triplet-based STDP can be used to classify complex patterns and it is also capable of reproducing several biological experiments including (i) those that consider higher order spike trains (e.g. triplets and quadruplets of spikes)~\cite{ref2,ref5,ref7}, and (ii) those that in addition to timing difference between pairs of spikes, bring the rate of spike-pairs into action of changing the synaptic weight~\cite{ref4}. However, to the best of our knowledge the link between SDSP and its ability of reproducing the outcomes of these biological experiments is not established or demonstrated. 

In this paper, we present a neuromorphic implementation, where a rate-based BCM rule emerges from a VLSI circuit implementation of triplet-based STDP. This circuit implements a time-based model of synaptic change where synaptic weights are altered using the timing difference between pre- and post-synaptic spikes; however, when periods of random pre- and post-synaptic activities are considered, then it can exhibit the behaviour of a rate-based model; in this current study BCM (a rate-based learning rule) is an emergent property.

The remainder of the paper is organized as follows. Section~\ref{sec:syn mod} provides a brief introduction to STDP (time-based model of synaptic plasticity) as well as BCM (rate-based model of synaptic plasticity) synaptic modification rules, and discusses how they are related. Section~\ref{sec:stdp vlsi} describes the circuit implementations of the proposed triplet-based STDP circuit alongside a previously proposed pair-based STDP circuit. Applied protocol and circuit experimental results are demonstrated in Section~\ref{sec:app prot}. Followed by a conclusion in section~\ref{conclusion}. 


\section{Synaptic modification rules} \label{sec:syn mod}

As already mentioned there are two major classes of synaptic modification rules. STDP rules modify synaptic weights according to the timing difference between occurrences of presynaptic/postsynaptic spikes~\cite{ref1}. One may consider just the timing of a pair of pre- and postsynaptic spikes~\cite{ref5}, while one may take the effect of a triplet/quadruplet or higher order sets of spike timing patterns into account~\cite{ref3,ref7}. Another important synaptic modification rule is the BCM rule which changes the syanptic strength according to the rate of pre- and post-synaptic action potentials. In the following, firstly a brief review of two known STDP models is provided, then the BCM learning rule is introduced, and finally the relation between STDP and the BCM is discussed~\cite{ref5}.


\subsection{Pair-based STDP} \label{sec:pair stdp}

Pair-based rule is the classical description of STDP, which has been widely used in various studies as well as many physical VLSI implementations~\cite{ref10,ref13,ref14,ref16}. Eq. 1 is a mathematical representation of the pair-based STDP rule~\cite{ref5}.

\begin{equation}\label{eq:stdp}
\Delta w = \begin{cases} \Delta w^+=A^+\exp(\frac{-\Delta t}{\tau_+}) & \mbox{if}~\Delta t>0 \\ \Delta w^-=-A^-\exp(\frac{\Delta t}{\tau_-}) & \mbox{if}~\Delta t \leq 0~, \end{cases}
\end{equation}
where $\Delta t=t_{\rm post}-t_{\rm pre}$ is the time difference between a single pair of pre- and post-synaptic spikes. 
According to this model, synaptic weight will be potentiated if a postsynaptic spike arrives in a specified time window ($\tau_+$) after the occurrence of a presynaptic spike. Analogously, depression will occur if a presynaptic spike takes place in a particular time window ($\tau_-$) after a postsynaptic spike. The amount of potentiation/depression will be determined as a function of the timing difference between pre- and post-synaptic spikes and their relevant amplitude parameters ($A^+$ and $A^-$).


\subsection{Triplet-based STDP} \label{sec:triplet stdp}

In this model of synaptic plasticity, synaptic weight changes based on the timing of a triplet combination of spikes (e.g. pre-post-pre, or a post-pre-post)~\cite{ref7}. The mathematical representation of this learning rule, as shown in \cite{ref7}, is given by

\begin{equation}\label{eq:stdptrip}
\Delta w = \begin{cases} \Delta w^+=\exp(\frac{-\Delta t_1}{\tau_+})\Big(A_2^+ +A_3^+\exp(\frac{-\Delta t_2}{\tau_y})\Big) \\ \Delta w^-=-\exp(\frac{\Delta t_1}{\tau_-})\Big(A_2^- +A_3^-\exp(\frac{-\Delta t_3}{\tau_x})\Big),\end{cases}
\end{equation}
where $\Delta w=\Delta w^+$ if $\ t=t_{\rm post}$ and $\Delta w=\Delta w^-$ if $\ t=t_{\rm pre}$. $A_2^+$, $A_2^-$, $A_3^+$ and $A_3^-$ are amplitude constants, $\Delta t_1=t_{\rm post}-t_{\rm pre}$, $\Delta t_2=t_{{\rm post}(n)}-t_{{\rm post}(n-1)}-\epsilon$ and $\Delta t_3=t_{{\rm pre}(n)}-t_{{\rm pre}(n-1)}-\epsilon$, are time difference between combinations of pre- and post-synaptic spikes. $\epsilon$ is a small positive constant which ensures that the weight update happens at the right time, and finally $\tau_-$, $\tau_+$, $\tau_x$ and $\tau_y$ are time constants \cite{ref7}.

In triplet-based STDP, potentiation can occur when a post-synaptic spike is generated. When this post-synaptic spike occurs, the synaptic weight will be increased by means of two different interactions. First is the interaction between current post-synaptic spike and its previous pre-synaptic spike by an amount proportional to $A_2^+$. Second is the interaction of the current post-synaptic spike and its preceding post-synaptic spike by an amount proportional to $A_3^+$. An analogous description holds for synaptic depression, i.e. depression is the result of a pre-synaptic pulse after a post-synaptic pulse (amplitude $A_2^-$), and also a pre-synaptic pulse after a pre-synaptic pulse (amplitude $A_3^-$). It is worth mentioning that, the first potentiation/depression term has a direct effect on its relative second potentiation/depression terms (see Eq. 2) that leads to nonlinearity in triplet-based STDP.

Previous studies have illustrated that classical pair-based STDP fails to reproduce the experimental outcomes involving higher order spike patterns such as spike triplets and quadruplets~\cite{ref3,ref7,froemke2006} and, furthermore, fails to account for the observed dependence on repetition frequency of pairs of spikes~\cite{ref4,ref7,froemke2006}. Triplet-based STDP was developed to resolve the shortcomings of the pair-based STDP~\cite{ref4,ref7}. 


\subsection{The BCM rule} \label{sec:bcm rule}

Unlike STDP which is spike-timing based learning rule, synaptic modifications resulting from the BCM rule depends on the rate of the pre- and post-synaptic spikes~\cite{ref9}. In fact, it depends linearly on the pre-synaptic, but nonlinearly on the post-synaptic spike rate. The mathematical model of the BCM learning rule has been demonstrated in different ways, but a general, yet simple form of this model is given by~\cite{ref7}

\begin{equation}\label{eq:bcm}
\frac{\Delta w}{\Delta t} = \rho_x \phi(\rho_y,\theta),
\end{equation}
where $\rho_x$ and $\rho_y$ are the pre-synaptic and post-synaptic spike rates and $\theta$ is a constant which represents some threshold ~\cite{ref7}. In addition, when $\phi(\rho_y<\theta,\theta)<0$ synaptic weights will be decreased (depression), and when $\phi(\rho_y>\theta,\theta)>0$, they will be increased (potentiation) and if $\phi(0,\theta)=0$, there will be no change in synaptic weight~\cite{ref7}.


\subsection{Relation of STDP to the BCM} \label{sec:relation}

According to the literature, the BCM rule can emerge from pair-based and triplet-based STDP rules. In 2003, Izhikevich and Desai~\cite{ref15} demonstrated that, the nearest-spike interaction \footnote{Nearest-spike model considers the interaction of a spike only with its two immediate succeeding and immediate preceding nearest neighbours.} version of pair-based STDP can replicate BCM behaviour. Furthermore, Pfister and Gerstner, in 2006~\cite{ref7} have reported, a triplet-based model of STDP that can also produce BCM behaviour, when long-time spike statistics are taken into account. According to~\cite{ref15}, under the assumption of Poissonian distribution of spike times for pre-synaptic and post-synaptic spike trains, nearest-spike pair-based STDP can give rise to the BCM rule; i.e. BCM emerges from nearest neighbour pair-based STDP; while all-to-all\footnote{All-to-all model considers the interaction of every single spike with all other spikes, not only with its nearest neighbours.} spike interaction cannot. Furthermore, based on~\cite{ref7}, if the pre-synaptic and post-synaptic spike trains in a triplet-based STDP model are Poissonian spike trains, then BCM learning is an emergent property of the model. In the next section, VLSI implementations of nearest-spike interaction for both pair-based and triplet-based STDP are presented. Then, in the following section, adopting a Poissonian protocol, these circuits are tested to verify the emergent relationship between (nearest-spike) STDP and BCM learning. It is worth mentioning that, during all experiments in this paper, only nearest spike interactions are considered.


\section{STDP VLSI Implementation} \label{sec:stdp vlsi}

\subsection{VLSI Implementation of Pair-based STDP} \label{sec:pair vlsi}

\begin{figure}
	\centering
	 \includegraphics[width=0.48\textwidth]{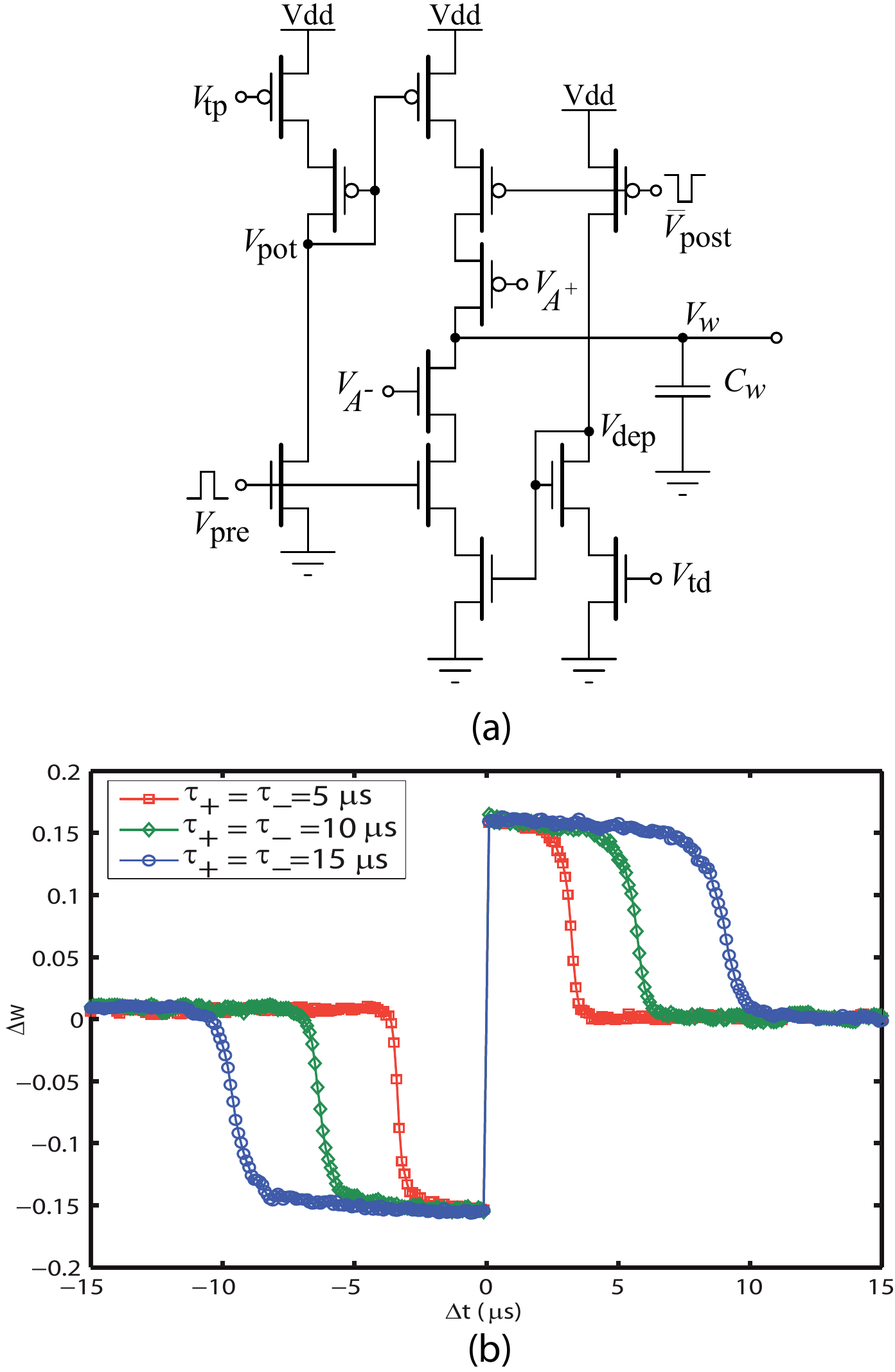}
	\caption{VLSI implementation of classical pair-based STDP. (a) Schematic circuit diagram of Indiveri~\emph{et al.} circuit~\cite{ref10}. (b) The learning window of the circuit based on our simulations.}\label{fig:1}
\end{figure}

There are several VLSI implementations of a pair-based STDP in the literature~\cite{ref10,ref13,ref14,ref16}, where the implementation by Indiveri~\emph{et al.}~\cite{ref10} was adopted, due to its low power and small area. Fig.~\ref{fig:1}(a) depicts the schematic circuit diagram for Indiveri's pair-based STDP circuit and Fig.~\ref{fig:1}(b) demonstrates its resulting temporal learning window for various $\tau_+$ and $\tau_-$ ($V_{\rm tp}$,$V_{\rm td}$) based on our simulations. In this circuit, the timing of pre- and post-synaptic spikes are used to induce weight changes across a load capacitor, $C_w$. When a presynaptic pulse, $V_{\rm pre}$, or a postsynaptic pulse ($\overline{V}_{\rm post}$) occurs, $V_{\rm pot}$ ($V_{\rm dep}$) will be set to zero ($V_{\rm dd}$). This $V_{\rm pot}$($V_{\rm dep}$) then changes linearly over time to reach $V_{\rm dd}$  (zero), and represents the required time constants $\tau_+$($\tau_-$). These time constants can be set by changing the gate voltage of the corresponding transistors, i.e. $V_{\rm tp}$ ($V_{\rm td}$). If a $V_{\rm pre}$ ($\overline{V}_{\rm post}$) pulse occurs during the time determined by its corresponding time constant, $\tau_-$ ($\tau_+$), the output capacitor will be discharged (charged) by a current that is proportional to the value of $V_{\rm dep}$ ($V_{\rm pot}$) at that time and also $V_{\rm A^-}$ ($V_{\rm A^+}$). For further details please refer to~\cite{ref10}.

\subsection{VLSI Implementation of Triplet-based STDP} \label{sec:triplet vlsi}

\begin{figure*}
	\centering
	 \includegraphics[width=0.65\textwidth]{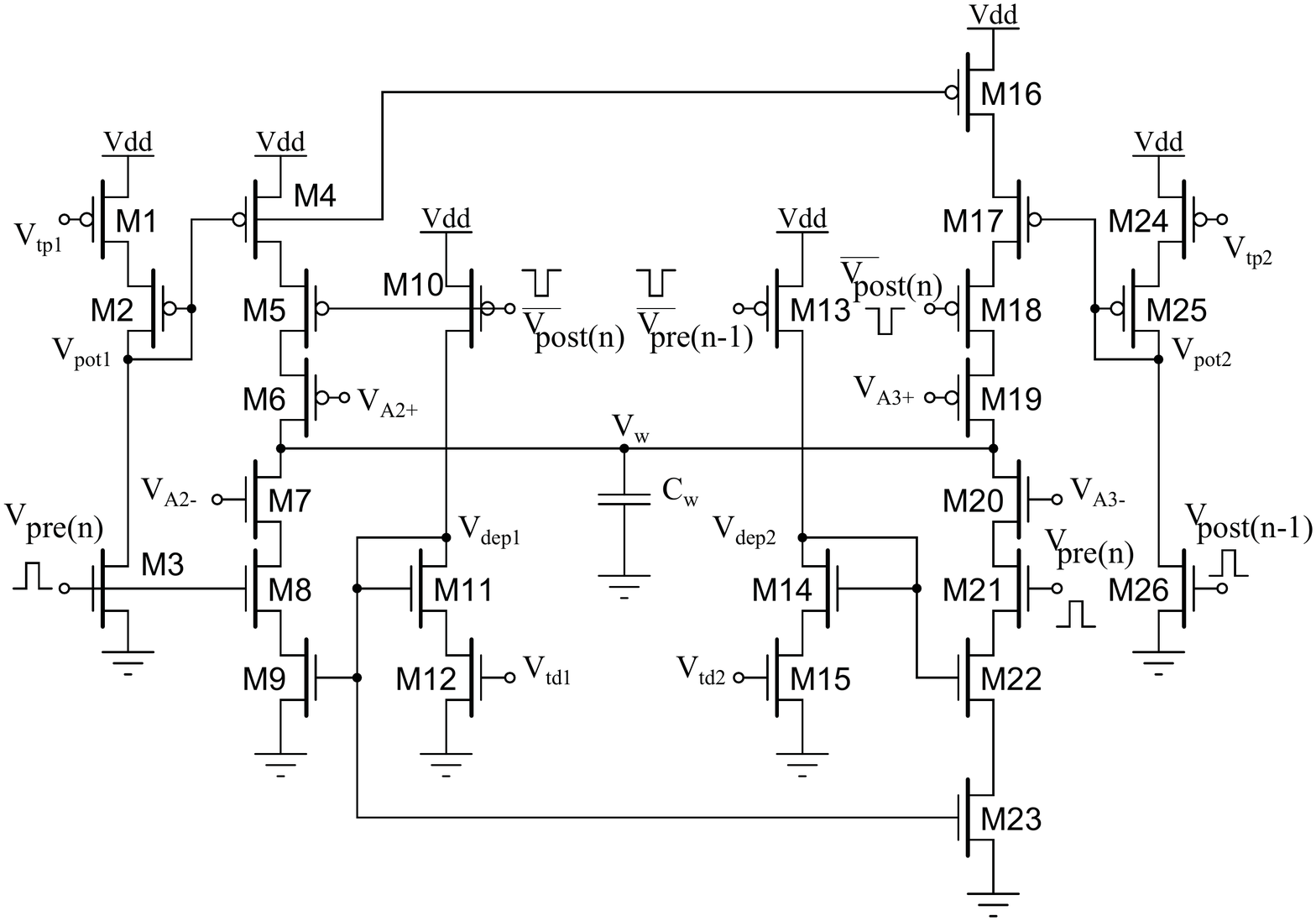}
	\caption{VLSI implementation of triplet-based STDP} \label{fig:2}
\end{figure*}

Unlike the pair-based model; synaptic change in the triplet model is induced from patterns consisting of three consecutive spikes, where the pattern consists of two pre-synaptic and one post-synaptic or two post-synaptic and one pre-synaptic spikes. The proposed triplet-based STDP circuit, which is configured to consider the spike triplet interactions in generating Long Term Potentiation (LTP), or Long Term Depression (LTD), is an extension to the pair-based circuit introduced in~\cite{ref10}. In the proposed triplet circuit, two more pulses, $V_{{\rm post}(n-1)}$ and $\overline{V}_{{\rm pre}(n-1)}$, are used in addition to $\overline{V}_{{\rm post}(n)}$ and $V_{{\rm pre}(n)}$, as shown in Fig.~\ref{fig:2}. 
      
This circuit operation can be described as follows: upon the arrival of a post-synaptic pulse, $\overline{V}_{{\rm post}(n)}$, the M5, M10 and M18 transistor switches turn on. Then M10 sets a depotentiating voltage $V_{\rm dep1}$ to $V_{\rm dd}$. This voltage then starts decaying linearly in time which can result in depression, if a pre-synaptic pulse, $V_{{\rm pre}(n)}$ arrives during the time $V_{\rm dep1}$ is decaying to zero ($\tau_-$ time constant). In this situation,~$C_w$ will be discharged through M7-M9 by a current that is limited by the M7 bias voltage ($V_{\rm A_2^-}$). In contrast to M10, which can result in depression after receiving a post-synaptic pulse, M5 and M18 can lead to two different potentiations. The first one can occur if M5 turns on during time constant of $V_{\rm pot1}$ ($\tau_+$). This potentiation will be through M4-M6 and is proportional to the bias voltage at M6 ($V_{\rm A_2^+}$). The second potentiation term can charge $C_w$ through M16-M19 and is proportional to $V_{\rm A_3^+}$ if M18 is on at the required time, i.e. when $V_{\rm pot1}$ and $V_{\rm pot2}$ have still kept M16 and M17 on. This is the term that distinguishes triplet from pair-based STDP, as there is no such term in pair-based STDP. Similarly, upon the arrival of a pre-synaptic pulse, $V_{{\rm pre}(n)}$, a potentiating voltage $V_{\rm pot1}$ is set to zero and starts increasing linearly in time which can result in potentiation when a $\overline{V}_{{\rm post}(n)}$ pulse arrives within the $\tau_+$ time constant. In addition, two possible depressions proportional to $A_2^-$ and $A_3^-$ can take place, if this pre-synaptic pulse is in the interval area of effect of $V_{{\rm dep1}}$ and $V_{{\rm dep2}}$, i.e. in $\tau_-$ and $\tau_x$ time constants. 

In addition to the property where BCM learning emerges from the triplet-based VLSI STDP circuit, this implementation is also capable of reproducing some other challenging biological experiments such as the affect of pairing frequency of spike pairs on weight change, as well as higher-order spike interaction consideration~\cite{ref7,rahimitriplet}. 


\section{Applied Protocol and Circuit Experimental Results} \label{sec:app prot}

\subsection{Poisson Protocol} \label{sec:pois prot}
In order to directly test whether BCM learning emerges from the two different STDP circuits, similar experimental protocol to those employed in~\cite{ref7,ref15} was adopted. Under this protocol, the pre-synaptic and post-synaptic spike trains are Poissonian spike trains, with the rate of $\rho_x$ for pre-synaptic, and $\rho_y$ for post-synaptic spikes. This means that the inter-spike intervals between spikes can be calculated as: $P_x(s)=\rho_x$exp$(-\rho_x s)$ for the pre-synaptic spikes and $P_y(s)=\rho_y$exp$(-\rho_y s)$ for the post-synaptic spikes. Using this protocol, the integral of the STDP temporal learning window function can be (directly) mapped to the BCM rule. As presented in~\cite{ref15}, synaptic weight alteration for a nearest-spike model of pair-based STDP per one presynaptic spike can be given by

\begin{equation}\label{eq:bcm stdp}
C(\rho_y)= \rho_y \left(\frac {A_+}{\tau_+^{\rm -1}+\rho_y}+\frac{A_-}{\tau_-^{\rm -1}+\rho_y}\right).
\end{equation}
 
Moreover, under Poissonian spike statistics, nearest-spike triplet-based STDP rule, as well as all-to-all triplet-based STDP, can both be mapped to BCM~\cite{ref7}. In this paper, only the nearest-spike model is considered. The total synaptic weight change induced by nearest-spike triplet STDP model, can be expressed by Eq.~\ref{eq:triplet bcm}~\cite{ref7}.

\begin{equation}\label{eq:triplet bcm}
\begin{split}
\left\langle\frac{dw}{dt}\right\rangle=-A_2^- \frac {\rho_x\rho_y}{\tau_-^{\rm -1}+\rho_y}-A_3^- \frac {\rho_x^2\rho_y}{(\tau_-^{\rm -1}+\rho_y)(\tau_x^{\rm -1}+\rho_x)}
\\ + A_2^+ \frac {\rho_x\rho_y}{\tau_+^{\rm -1}+\rho_x}+A_3^+ \frac {\rho_x\rho_y^2}{(\tau_+^{\rm -1}+\rho_x)(\tau_y^{\rm -1}+\rho_y)}.
\end{split}
\end{equation}

\begin{figure}
	\centering
	 \includegraphics[width=0.48\textwidth]{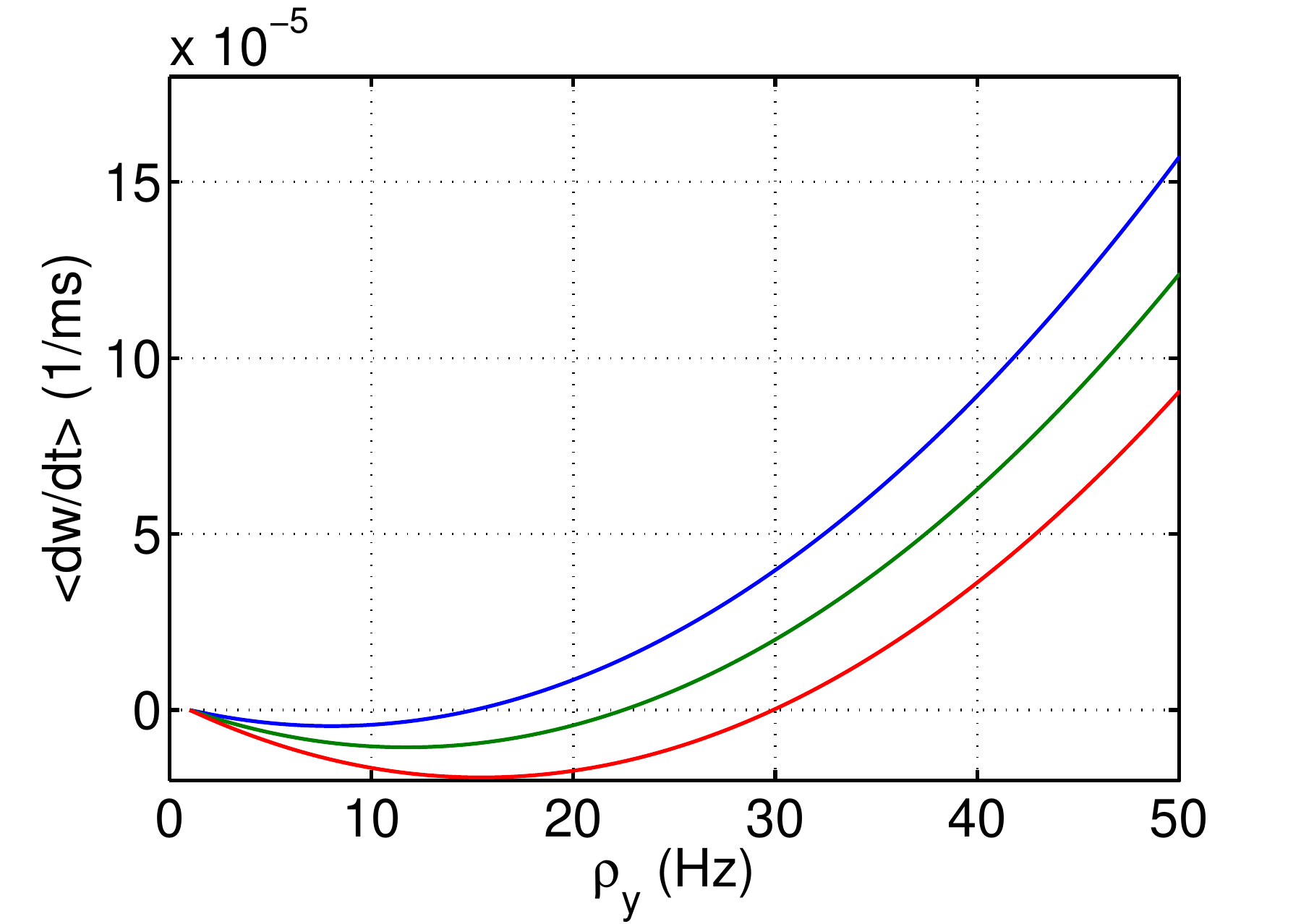}
	\caption{Matlab simulation results of the all-to-all triplet-based STDP~\cite{ref7} as a BCM rule. Three different curves show synaptic weight changes according to three different synaptic modification thresholds which demonstrate the points where LTD changes to LTP. The threshold is adjustable using the STDP rule parameters (please refer to the text).} \label{bcmout}
\end{figure}

Based on the rate-based BCM rule, synaptic weight change is linearly dependent on $\rho_x$ and nonlinearly depends on $\rho_y$ (see Eq.~\ref{eq:bcm}). In order to satisfy this condition in Eq.~\ref{eq:triplet bcm}, $A_3^-$ must be equal to zero and also $\rho_x\ll\tau_+^{\rm -1}$. This is a minimal case of the triplet-based STDP model (please refer to~\cite{ref7}). Also, based on the BCM learning rule definition, the synaptic weight modification threshold is a function of postsynaptic activity, i.e. $\theta=\alpha\left\langle\rho_y^p\right\rangle$ where $p>1$. For triplet-based STDP, consider the case where all-to-all interactions between triplets of pre- and post-synaptic spikes; it is possible to redefine $A_2^-$, $A_2^+$ and $A_3^+$ in a way that the threshold be dependent on the postsynaptic firing rate, $\rho_y^p$. However, in the nearest-spike model it is not possible to change these parameters in a way that $\theta=\alpha\left\langle\rho_y^p\right\rangle$. Although the triplet-based nearest-spike STDP model cannot fully satisfy the second condition of a BCM learning rule (the dependency of threshold on $\rho_y$), it can elicit the properties of BCM for a limited range of frequencies. Matlab simulation results (Fig.~\ref{bcmout}) show how the threshold is modulated by controllable amplitude parameters ($A_2^-$ , $A_2^+$ and $A_3^+$) for all-to-all interaction. (Please refer to the text and also supplementary materials of~\cite{ref7}).

\subsection{Circuit Experimental Results} \label{circuit exp}
\begin{figure}
	\centering
	 \includegraphics[width=0.5\textwidth]{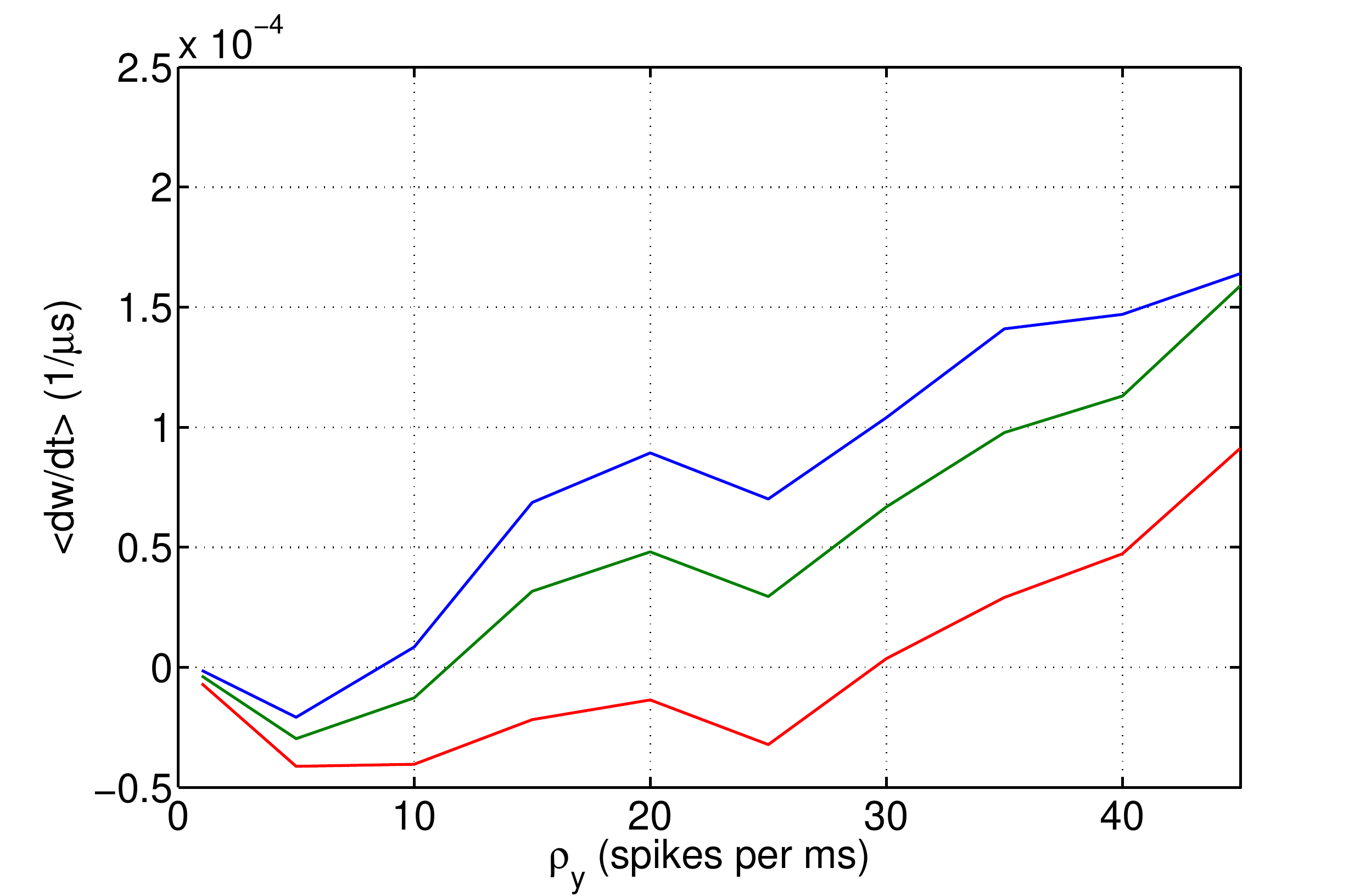}
	\caption{Simulation results of the nearest-neighbour pair-based STDP (Fig.~\ref{fig:1}) as a BCM rule. Three different curves show synaptic weight changes according to three different synaptic modification thresholds which demonstrate the points where LTD changes to LTP. The threshold is adjustable using the STDP rule parameters (please refer to the text).} \label{fig:3}
\end{figure}

Fig.~\ref{fig:3} depicts the simulation results for the pair-based STDP circuit (Fig.~\ref{fig:1}), under the (earlier stated) Poisson protocol. In this circuit, $\overline{V}_{\rm post}$ and $V_{\rm pre}$ are Poissonian spike trains where the firing rate of pre- and post-synaptic spikes are $\rho_y$ and $\rho_x$, respectively. For the above  mentioned circuit, during the simulations, a BCM-like behaviour emerges from nearest-spike pair-based STDP as demonstrated in Fig.~\ref{fig:3}. In this circuit the modification threshold can be calculated by the following~\cite{ref15},

\begin{equation}\label{eq:threshold}
\theta=-\frac {A_+/\tau_{-} + A_-/\tau_+}{A_+ + A_-}.
\end{equation}

For producing the three different waveforms shown in Fig.~\ref{fig:3}, three different thresholds have been used by keeping $V_{\rm A_-}$, $V_{\rm tp}$ and $V_{\rm td}$ fixed and adjusting $V_{\rm A_+}$.

\begin{figure}
	\centering
	 \includegraphics[width=0.5\textwidth]{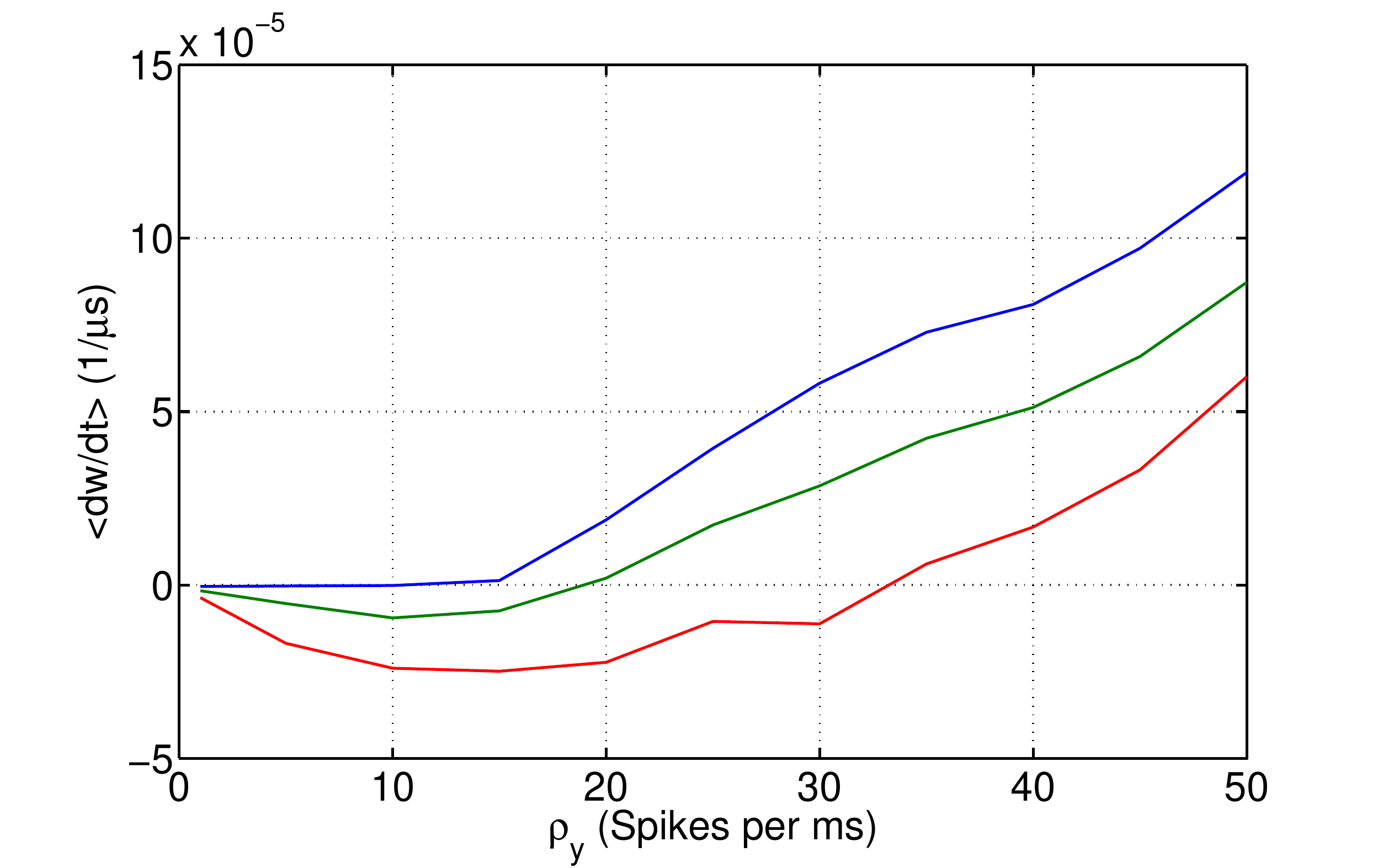}
	\caption{Simulation results of the nearest-neighbour triplet-based STDP (Fig.~\ref{fig:2}) as a BCM rule. Three different curves show synaptic weight changes according to three different synaptic modification thresholds which demonstrate the points where LTD changes to LTP. The threshold is adjustable using the STDP rule parameters (please refer to the text).} \label{fig:4}
\end{figure}

As already mentioned, besides pair-based STDP, BCM behaviour can also be replicated by triplet-based model of STDP~\cite{ref7}. Fig.~\ref{fig:2} demonstrates the proposed VLSI implementation of the nearest-spike triplet-based STDP rule. In order to take the BCM characteristics out of this STDP circuit, either $V_{\rm A_3^-}$ must be set to zero, or the second depression part of the circuit on the bottom right part of the circuit (M13-M15 and M20-M23) can be eliminated. This will be a minimal triplet-based STDP circuit which satisfies the first criteria of the BCM rule. In this circuit, $V_{\rm post(n-1)}$, $\overline{V}_{\rm post(n)}$ and $V_{\rm pre(n)}$ are Poissonian spike trains where $\rho_y$, $\rho_y$ and $\rho_x$ denote their firing rates, respectively. Fig.~\ref{fig:4} depicts the simulation results of this circuit under the Poisson protocol. Three different curves are presented which display three different weight modification thresholds. In the BCM rule, these thresholds are related to the post-synaptic firing rate, $\rho_y$. Based on~\cite{ref7}, the modification threshold changes for the case when all-to-all spike interactions  are considered and can be expressed as Eq.~\ref{eq:trip thr},

\begin{equation}\label{eq:trip thr}
\theta=\left\langle\rho_y^p\right\rangle \frac{(A_2^- \tau_- A_2^+ \tau_+)}{(\rho_0^p A_3^+ \tau_+ \tau_y)},
\end{equation}
where $\left\langle\rho_y^p\right\rangle$ is the expectation over the statistics of the $p^{\rm th}$ power of the post-synaptic firing rate, $\rho_0^p=\left\langle\rho_y^p\right\rangle$ for large time constants (10 min or more). However, for the nearest-spike model which is the case of the proposed circuit, it is not possible to derive an equation for the modification threshold based on $\overline{\rho_y^p}$, but for postsynaptic firing rate up to 50, a similar behaviour to what Eq.~\ref{eq:trip thr} presents is inferable from the simulation results (supplementary materials of~\cite{ref7}). 

It should be mentioned that, the simulation results are accelerated 1000 times when compared to real biological time. This means that for the proposed VLSI circuit, 1~ms of time used in the circuit represents one second of biological time. Put simply, in Fig.~\ref {fig:3}, $\rho_y$ is the average firing rate of the post-synaptic neuron is given by spikes per millisecond rather than spikes per second (here 1 ms of circuit time is equivalent to 1 second). This accelerated approach has been used in previous studies of implementing STDP with VLSI such as~\cite{ref12,ref16}.

The current circuit implementation in its present form is sensitive to variations in bias voltages, as it is the case with the original circuit by Indiveri~\emph{et al.}~\cite{ref10}. Hence, in the presented design, each bias voltage can be provided using a diode-connected MOS device in series with a current source, resulting in a current mirror structure. As the current values to achieve the needed biases are very low in the sub-nanoampere (nA) levels range, some of these transistors will be forced to operate in the subthreshold regime of operation. This approach will result in a more robust approach for bias voltage setting instead of direct biasing using fixed voltage sources. Furthermore, these current sources can be combined with digital trimming techniques to elevate some of these short comings due to process variations and allow for bias voltage adjustment even after circuit fabrication.

To date, there have been two different proposed VLSI circuits which implement the BCM rule~\cite{ref12,meng2011iono,rachmuth2011biophysically}. To the best of our knowledge, just the VLSI circuit in~\cite{ref12} has the same capabilities as the proposed triplet-based STDP VLSI circuit presented in this paper. That circuit implements a recently introduced BCM-like model~\cite{ref17}, which is capable of replicating pairing frequency effects and triplet spike effects (quadruplet experiments have not been shown in this paper), and of course BCM behaviour. In comparison to the proposed circuit presented in this paper, the circuit proposed in~\cite{ref12} contains more devices, hence larger silicon area when implemented and more prone to process variation. The neuron architecture also needs to be changed in a way that is compatible to the learning algorithm, while our VLSI circuit simply acts as a plastic synapse, which can be used to connect to other sets of (neuromorphic) neurons of choice. Furthermore, the results from our proposed VLSI implementation can reproduce the outcomes of previously published BCM experiments \cite{clothiaux1991,kirkwood1996,ref9}.

 
\section{Conclusion} \label{conclusion}
A new VLSI implementation, based upon spike-timing leading to the emergence of the BCM rule was presented. The proposed implementation is based upon a triplet-based nearest-spike STDP model. This model can replicate the same BCM behaviour as previous studies~\cite{ref12,meng2011iono}, as well as other important synaptic plasticity experiments, including the effects of pairing frequency of spike pairs on weight change and higher-order spike interactions (triplet and quadruplet). It was also shown that the VLSI implementation of classical pair-based STDP, which is based on the interaction between pairs of pre- and post-synaptic spikes rather than the triplets of spikes can also reproduce similar behaviour as the BCM rule. The results of this study can be a powerful addition to classical pair-based STDP, which is known not to be capable of mimicking other plasticity based experimental phenomena. The presented design can play an important role in hardware implementation of spiking-based neural networks.

\section*{Acknowledgment}
The support of the Australian Research Council (ARC) is gratefully acknowledged.

\bibliographystyle{ieeetran}
\bibliography{refs}

\end{document}